\documentclass[aps,prd,preprint,groupedaddress,showpacs]{revtex4}

\usepackage{epsfig}
\usepackage{graphics}
\usepackage{slashed}
\usepackage{color}
\begin{document}

\leftmargin -2cm
\def\choosen{\atopwithdelims..}
\leftmargin -2cm
\def\choosen{\atopwithdelims..}

~~\\
 DESY~14--199 \hfill ISSN 0418-9833
\\October 2014

\vspace{2cm}
\title{Open charm production in the parton Reggeization approach: from Tevatron to LHC}
 \author{\firstname{M.A. }\surname{Nefedov}}
\email{nefedovma@gmail.com} \author{\firstname{A.V.}
\surname{Karpishkov}} \email{karpishkov@rambler.ru}
  \affiliation{Samara State University, Ac.\ Pavlov, 1, 443011 Samara,
Russia}
\author{\firstname{V.A.}
\surname{Saleev}} \email{saleev@samsu.ru} \affiliation{Samara State University, Ac.\ Pavlov, 1, 443011 Samara,
Russia}\affiliation{Samara State Aerospace University, Moscow
Highway, 34, 443086, Samara, Russia}
\author{\firstname{A.V.}
\surname{Shipilova}} \email{alexshipilova@samsu.ru}
\affiliation{Samara State University, Ac.\ Pavlov, 1, 443011 Samara,
Russia}\affiliation{{II.} Institut f\"ur Theoretische Physik,
Universit\" at Hamburg, Luruper Chaussee 149, 22761 Hamburg,
Germany}

\begin{abstract}
 We study the inclusive hadroproduction of $D^0$, $D^+$, $D^{\star
+}$, and $D_s^+$ mesons at leading order in the parton Reggeization
approach endowed with universal fragmentation functions fitted to
$e^+e^-$ annihilation data from CERN LEP1. We have described
$D$-meson transverse momentum distributions measured in the central
region of rapidity by the CDF Collaboration at Tevatron ($|y|<1$)
and ALICE Collaboration at LHC ($|y|<0.5$) within uncertainties and
without free parameters, using Kimber-Martin-Ryskin unintegrated
gluon distribution function in a proton. The forward $D$-meson
production ($|y|>2.0$) measured by the LHCb Collaboration also has
been studied and expected disagreement between our theoretical
predictions and data has been obtained.
\end{abstract}
\pacs{12.38.-t,12.40.Nn,13.85.Ni,14.40.Gx}
 \maketitle

\section{Introduction}

The study of the open charm production in the high energy hadronic
collisions is considered as a test of general applicability of
perturbative quantum chromodynamics (QCD). In the process of charmed
meson production one has $\mu\ge m$, where $\mu$ is the typical
energy scale of the hard interaction, $m$ is the charm quark mass,
and $\alpha_S(\mu)\ll 1$. Nevertheless, this study is also our
potential for the observation of a new dynamical regime of
perturbative QCD, namely the high-energy \textit{Regge limit}, which
is characterized by the following condition $\sqrt{S}\gg
\mu\gg\Lambda_{QCD}$, where $\sqrt{S}$ is the invariant collision
energy, and $\Lambda_{QCD}$ is the asymptotic scale parameter of
QCD. In this limit a new small parameter $x\sim\mu/\sqrt S$ appears.

The small-$x$ effects cause the distinction of
the perturbative corrections relative for different processes and
different regions of phase space. At first, the higher-order
corrections for the production of heavy final states, such as Higgs
bosons, top-quark pairs, dijets with large invariant masses, or
Drell-Yan pairs, by initial-state partons with relatively large
momentum fractions $x\sim 0.1$ are dominated by soft and collinear
gluons and may increase the cross sections up to a factor 2. By
contrast, relatively light final states, such as
small-transverse-momentum heavy quarkonia, single jets, prompt
photons, or dijets with small invariant masses, are produced by the
fusion of partons with small values of $x$, typically $x\sim
10^{-3}$ because of the large values of $\sqrt{S}$. Radiative
corrections to such processes are dominated by the production of
additional hard jets. The only way to treat such processes in the
conventional collinear parton model (CPM) is to calculate
higher-order corrections in the strong coupling constant
$\alpha_S=g_S^2/4\pi$, which could be a challenging task for some
processes even at the next-to-leading order (NLO) level. To overcome
this difficulty and take into account a sizable part of the
higher-order corrections in the small-$x$ regime, the
$k_T$-factorization framework, was introduced \cite{KTCollins,KTGribov,KTCatani}.

Recently the ALICE Collaboration measured the differential cross
sections $d\sigma /dp_T$ for the inclusive production of $D^0$,
$D^+$, $D^{\star +}$, and $D_s^+$ mesons
\cite{ALICE276,ALICE7,ALICE7Ds} in proton-proton collisions at the
CERN LHC ($\sqrt{S}=2.76; 7$~TeV) as functions of $D$-meson
transverse momentum ($p_T$) in the central rapidity region,
$|y|<0.5$. These measurements extend the CDF Collaboration data
\cite{CDFdmeson} obtained earlier in proton-antiproton collisions at
the Fermilab Tevatron at the $|y|<1.0$ and $\sqrt{S}=1.96$ TeV. The
production of $D$-mesons in the forward rapidity region of
$2.0<y<4.5$ was investigated at the LHC by LHCb Collaboration and
the data in the form of $d\sigma /dp_T$ were presented for the
different intervals of rapidity~\cite{forward}.

These data have been studied in the next-to-leading order (NLO) in
the collinear parton model of QCD within the two approaches: the
general-mass variable-flavor-number (GM-VFN) scheme~\cite{GMVFN},
and the so-called fixed order scheme improved with next-to-leading
logarithms (FONLL scheme)~\cite{FONLL}. In the former one, realized
in the Refs.~\cite{KniehlKramer, KKSS, KKSS2012}, the large
fragmentation logarithms dominating at $p_T>>m$ are resummed through
the evolution of the fragmentation functions (FFs), satisfying the
Dokshitzer-Gribov-Lipatov-Altarelli-Parisi (DGLAP)~\cite{DGLAP}
evolution equations. At the same time, the full dependence on the
charm-quark mass in the hard-scattering cross section is retained to
describe consistently $p_T\sim m$ region. The $D$-meson FFs were
extracted both at leading and next-to-leading order in the GM-VFN
scheme from the fit of $e^+e^-$ data taken by the OPAL Collaboration
at CERN LEP1~\cite{LEP1}. Opposite, in the FONLL approach, the NLO
$D$-meson production cross sections are calculated with a
non-perturbative $c$-quark FF, that is not a subject to
DGLAP~\cite{DGLAP} evolution. The FONLL scheme was implemented in
the Refs.~\cite{Nason,newNLO} and its main ingredients are the
following: the NLO fixed order calculation (FO) with resummation of
large transverse momentum logarithms at the next-to-leading level
(NLL) for heavy quark production. For the consistency of the
calculation, the NLL formalism should be used to extract the
nonperturbative FFs from $e^+e^-$ data, and in the
Refs.~\cite{Nason,newNLO} the scheme of calculation of heavy quark
cross section and extraction
 of the nonperturbative FFs are directly connected and must be used only together.

The overall agreement of data and calculations obtained in
Refs.~\cite{Nason,newNLO,KniehlKramer,KKSS,KKSS2012} is good, the $D-$meson spectra
measured by the CDF Collaboration at the Fermilab Tevatron and ALICE and LHCb
Collaborations at the LHC are described within experimental uncertainties.


The aim of the present work is to study the $D$-meson production at Fermilab Tevatron and CERN LHC in the
 framework of high-energy factorization scheme, namely $k_T$-factorization framework~\cite{KTCollins}
 endowed with the fully gauge-invariant amplitudes with \textit{Reggeized} gluons in the initial state.
 This combination we will call the Parton Reggeization Approach (PRA) everywhere below.

 The study of $D$-meson fragmentation production in terms of $k_T$-factorization~\cite{KTCollins,KTGribov,KTCatani}
 was performed also previously
  in the recent work~\cite{newKT}, with off-shell initial gluons and using the formalism of transverse-momentum
  dependent parton distributions, whereas the first results in this scheme were obtained for the $D_0$ production
  at Tevatron Run I~\cite{TeryaevBB}. The resulting curve in Ref.~\cite{newKT} describes the ALICE experimental data~\cite{ALICE7}
  by its upper limit of theoretical uncertainty. We suppose PRA to be more theoretically consistent  than previous studies in
   $k_T$-factorization,
  being not a recipe but based on a gauge invariant effective theory for the processes in quasi-multi-Regge kinematics (QMRK) in QCD.
   Therefore it preserves the gauge invariance of high-energy particle
  production amplitudes and allows a consistent continuation towards the NLO calculations.

Recently, PRA was successfully applied for the analysis of inclusive
production of single jet \cite{KSSY}, pair of jets \cite{NSSjets},
prompt-photon \cite{tevatronY,heraY}, photon plus jet \cite{KNS14},
Drell-Yan lepton pairs \cite{NNS_DY}, bottom-flavored jets
\cite{bbTEV,bbLHC}, charmonium and bottomonium production
\cite{KniehlSaleevVasin1,KniehlSaleevVasin2,PRD2003,NSS_charm,NSS_bot}
at the Tevatron and LHC. These studies  have demonstrated the
advantages of the high-energy factorization scheme based on PRA in
the descriptions of data comparing to the collinear parton model
calculations.

This paper is organized as follows.
In Sec.~\ref{sec:two} we present basic formalism of our
calculations, the PRA and the fragmentation model. In
Sec.~\ref{sec:three} our results are presented in comparison with the
experimental data and discussed. In Sec.~\ref{sec:four} we summarize our conclusions.

 \boldmath
\section{Basic Formalism}
\unboldmath
\label{sec:two}

The phenomenology of strong interactions at high energies exhibits a
dominant role of gluon fusion into heavy quark and antiquark pair in
heavy meson production. As it was shown in Ref.~\cite{KKSS}, a
significant part of $D$-meson production cross section comes from
gluon and $c$-quark fragmentation into $D$-meson, and the light
quark fragmentation turns out to be negligible. Following this, in our
study we will consider the $c$-quark and gluon fragmentation into different
$D$-mesons only.

In hadron collisions the cross sections of processes with a hard
scale $\mu$ can be represented as a convolution of scale-dependent
parton (quark or gluon) distributions and squared hard parton
scattering amplitude. These distributions correspond to the density
of partons in the proton with longitudinal momentum fraction $x$
integrated over transverse momentum up to $k_T=\mu$. Their evolution
from some scale $\mu_0$, which controls a non-perturbative regime,
to the typical scale $\mu$ is described by DGLAP~\cite{DGLAP}
evolution equations which allow to sum large logarithms of type
$\log(\mu^2/\Lambda_{QCD}^2)$ (collinear logarithms). The typical
scale $\mu$ of the hard-scattering processes is usually of order of
the transverse mass $m_T=\sqrt{m^2+|{\bf p}_T|^2}$ of the produced
particle (or hadron jet) with (invariant) mass $m$ and transverse
two-momentum ${\bf p}_T$. With increasing energy, when the ratio of
$x \sim \mu/\sqrt S$ becomes small, the new large logarithms
$\log(1/x)$, soft logarithms, are to appear and can become even more
important than the collinear ones. These logarithms present both in
parton distributions and in partonic cross sections and can be
resummed by the Balitsky-Fadin-Kuraev-Lipatov (BFKL)
approach~\cite{BFKL}.  The approach gives the description of QCD
scattering amplitudes in the region of large $S$ and fixed momentum
transfer $t$, $S \gg |t|$ (Regge region), with various color states
in the $t$-channel.
 Entering this region requires us to reduce approximations to keep the true kinematics of the process.
 It becomes possible introducing the unintegrated over transverse momenta parton distribution functions
 (UPDFs) $\Phi(x,t,\mu^2)$, which depend on parton transverse momentum
${\bf q}_T$ while its virtuality is $t=-|{\bf q}_T|^2$.  The UPDFs
are defined to be related with collinear ones through the equation:
\begin{eqnarray}
xG(x,\mu^2)=\int^{\mu^2}dt \Phi(x,t,\mu^2).
\end{eqnarray}
The UPDFs satisfy the BFKL evolution
equation \cite{BFKL} which is suited to resum soft logarithms and appear in the BFKL approach as a particular result in
the study of analytical properties of the forward scattering amplitude.
The basis of the BFKL approach is the gluon Reggeization \cite{gRegge}, as at small $x$ the gluons are the dominant partons.

The gluon Reggeization appears considering special types of
kinematics of processes at high-energies. At large $\sqrt S$ the
dominant contributions to cross sections of QCD processes gives
multi-Regge kinematics (MRK). MRK is the kinematics where all
particles have limited (not growing with $\sqrt S$) transverse
momenta and are combined into jets with limited invariant mass of
each jet and large (growing with $\sqrt S$) invariant masses of any
pair of the jets. At leading logarithmic approximation of the BFKL
approach (LLA), where the logarithms of type $(\alpha_s\log(1/x))^n$
are resummed, only gluons can be produced and each jet is actually a
gluon. At next-to-leading logarithmic approximation (NLA) the terms
of $\alpha_s(\alpha_s\log(1/x))^n$
 are collected and a jet can contain a couple of partons (two
gluons or quark-antiquark pair). Such kinematics is called quasi
multi-Regge kinematics. Despite of a great number of
contributing Feynman diagrams it turns out that at the Born level in
the MRK amplitudes acquire a simple factorized form. Moreover,
radiative corrections to these amplitudes do not destroy this form,
and their energy dependence is given by Regge factors
$s_i^{\omega(q_i)}$, where $s_i$ are invariant masses of couples of
neighboring jets and $\omega(q_i)$ can be interpreted as a shift of
gluon spin from unity, dependent from momentum transfer $q$. This
phenomenon is called gluon Reggeization.

Due to the Reggeization of quarks and gluons, an important role is
dedicated to the vertices of Reggeon-particle interactions. In
particular, these vertices are necessary for the determination of
the BFKL kernel. To define them we can notice the two ways: the
"classical" BFKL method~\cite{FadinFiore} is based on analyticity
and unitarity of particle production amplitudes and  the properties
of the integrals corresponding to the Feynman diagrams with two
particles in the $t$-channel has been developed. Alternatively, they
can be straightforwardly derived from the non-Abelian
gauge-invariant effective action for the interactions of the
Reggeized partons with the usual QCD partons, which was firstly
introduced in Ref.~\cite{KTLipatov} for Reggeized gluons only, and
then extended by inclusion of Reggeized quark fields in the
Ref.~\cite{LipVyaz}. The full set of the induced and effective
vertices together with Feynman rules one can find in
Refs.~\cite{LipVyaz,KTAntonov}.

Recently, an alternative method to obtain the gauge-invariant $2\to n$ amplitudes with off-shell initial-state partons, which is mathematically equivalent to the PRA, was proposed in Ref.~\cite{Kutak}. These $2\to n$ amplitudes are extracted by using the spinor-helicity representation with complex momenta from the auxiliary $2\to n+2$ scattering processes which are constructed to include the $2 \to n$ scattering processes under consideration. This method is more suitable for the implementation in automatic matrix-element generators, but for our study the use of Reggeized quarks and gluons is found to be simpler.

As we mentioned above, we will consider the $D$-meson production by only the $c$-quark and gluon fragmentation.
The lowest order in $\alpha_S$ parton subprocesses of PRA in which gluon or $c$-quark are produced are the following:
a gluon production
via two Reggeized gluon fusion
\begin{eqnarray}
\mathcal {R} + \mathcal {R} \to g,\label{eq:RRg}
\end{eqnarray}
and the corresponding quark-antiquark pair production
\begin{eqnarray}
\mathcal {R} + \mathcal {R} \to c + \bar c, \label{eq:RRQQ}
\end{eqnarray}
where $\mathcal {R}$ are the Reggeized gluons.

\begin{figure}[h]
\begin{center}
\includegraphics[width=.9\textwidth, clip=]{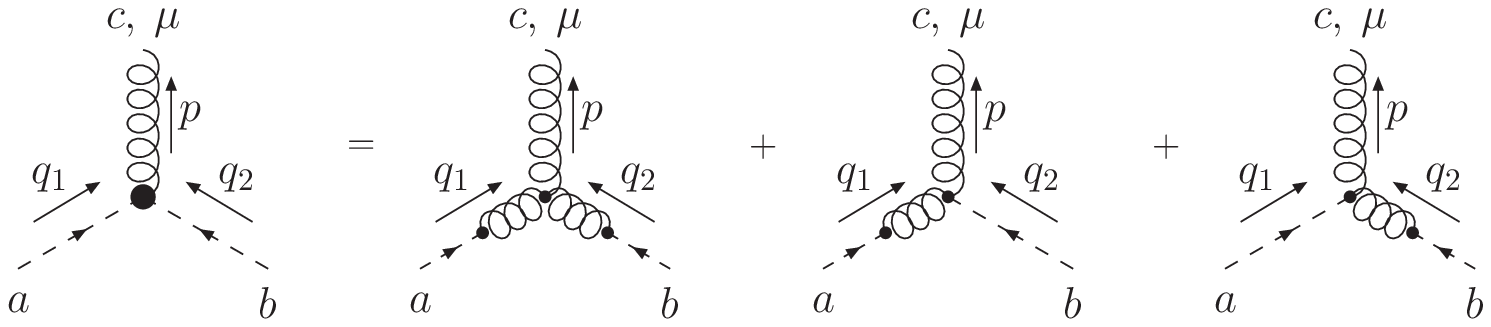}
\caption{Feynman diagrams for the subprocess (\ref{eq:RRg}). \label{fig:RRg}}
\end{center}
\end{figure}

\begin{figure}[h]
\begin{center}
\includegraphics[width=.9\textwidth, clip=]{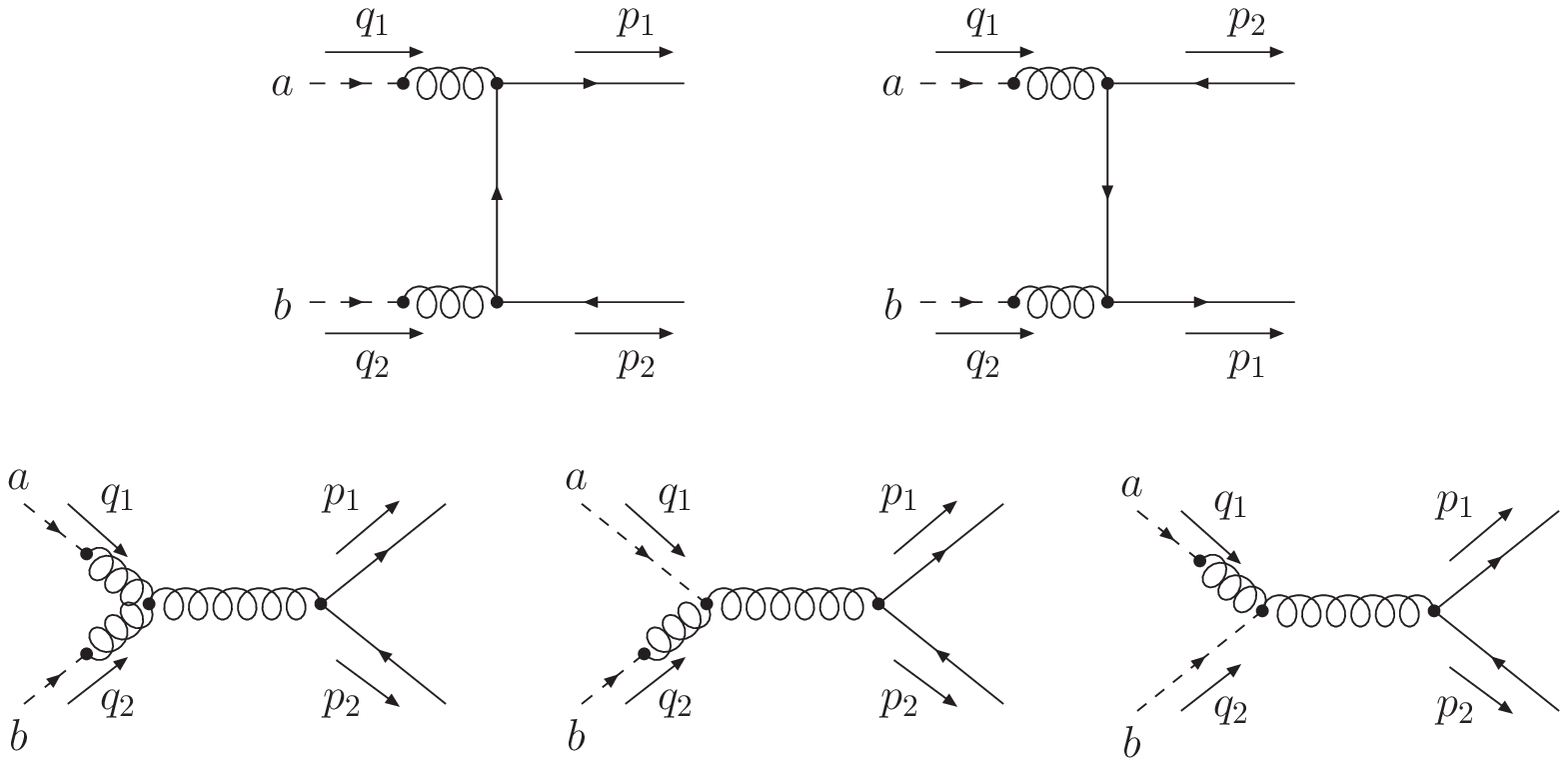}
\caption{Feynman diagrams for the subprocess (\ref{eq:RRQQ}).
\label{fig:RRQQ}}
\end{center}
\end{figure}

According to the prescription of Ref.~\cite{KTAntonov}, the
amplitudes of relevant processes (\ref{eq:RRg}) and (\ref{eq:RRQQ})
can be obtained from the Feynman diagrams depicted in
Figs.~\ref{fig:RRg} and \ref{fig:RRQQ}, where the dashed lines
represent the Reggeized gluons. Of course, the last three Feynman
diagrams in Fig.~\ref{fig:RRQQ} can be combined into the effective
particle-Reggeon-Reggeon  (PRR) vertex~\cite{KTAntonov}.

Let us define four-vectors $(n^-)^\mu=P_1^\mu/E_1$ and
 $(n^+)^\mu=P_2^\mu/E_2$, where $P_{1,2}^\mu$ are the four-momenta of
the colliding protons, and $E_{1,2}$ are their energies. We have
$(n^\pm)^2=0$, $n^+\cdot n^-=2$, and $S=(P_1+P_2)^2=4E_1E_2$. For
any four-momentum $k^\mu$, we define $k^\pm=k\cdot n^\pm$.  The
four-momenta of the Reggeized gluons can be represented as
\begin{eqnarray}
&&q_1^\mu = \frac{q_1^+}{2}(n^-)^\mu+q_{1T}^\mu\mbox{,}\nonumber\\
&&q_2^\mu = \frac{q_2^-}{2}(n^+)^\mu+q_{2T}^\mu\mbox{,}
\end{eqnarray}
where $q_{T}=(0,{\bf q}_{T},0)$
The amplitude of gluon production in fusion of two Reggeized gluons
can be presented as scalar product of Fadin-Kuraev-Lipatov effective
PRR vertex $C_{\mathcal{RR}}^{g,\mu}(q_1,q_2)$ and polarization
four-vector of final gluon $\varepsilon_\mu(p)$:
\begin{equation}
{\cal M}(\mathcal{R}+\mathcal{R}\to
g)=C_{\mathcal{RR}}^{g,\mu}(q_1,q_2)\varepsilon_\mu(p),
\end{equation}
where
\begin{eqnarray}
C_{\mathcal{RR}}^{g,\mu}(q_1,q_2)&=&-\sqrt{4\pi\alpha_s}f^{abc}
\frac{q_1^+q_2^-}{2\sqrt{t_1t_2}} \left[\left(q_1-q_2\right)^\mu+
\frac{(n^+)^\mu}{q_1^+}\left(q_2^2+q_1^+q_2^- \right)\right.\nonumber\\
&-&\left.\frac{(n^-)^\mu}{q_2^-}\left(q_1^2+q_1^+q_2^-\right)\right],
\label{amp:RRg}
\end{eqnarray}
$a$ and $b$ are the color indices of the Reggeized gluons with
incoming four-momenta $q_1$ and $q_2$, and $f^{abc}$ with
$a=1,...,N_c^2-1$ is the antisymmetric structure constants of color
gauge group $SU_C(3)$. The squared amplitude of the partonic
subprocess $\mathcal{R}+\mathcal{R}\to g$ is straightforwardly found
from Eq.~(\ref{amp:RRg}) to be
\begin{equation}
\overline{|{\cal M}(\mathcal{R}+\mathcal{R}\to g)|^2}=\frac{3}{2}\pi
\alpha_s \mathbf{p}_T^2. \label{sqamp:RRg}
\end{equation}

The amplitude of the process (\ref{eq:RRQQ}) can be presented in a
same way, as a sum of three terms ${\cal
M}(\mathcal{R}+\mathcal{R}\to c+\bar c)={\mathcal M}_1+{\mathcal
M}_2+{\mathcal M}_3$:
\begin{eqnarray}
&&{\mathcal M}_1=-i  \pi \alpha_s \frac{q_1^+ q_2^-}{\sqrt{t_1 t_2}}
T^aT^b \bar U(p_1) \gamma^\alpha\frac{\hat p_1-\hat
q_1}{(p_1-q_1)^2}\gamma^\beta V(p_2)(n^+)^\alpha(n^-)^\beta,
\nonumber\\
&&{\mathcal M}_2=-i \pi \alpha_s \frac{q_1^+ q_2^-}{\sqrt{t_1 t_2}}
T^bT^a\bar U(p_1) \gamma^\beta\frac{\hat p_1-\hat
q_2}{(p_1-q_2)^2}\gamma^\alpha
V(p_2)(n^+)^\alpha(n^-)^\beta,\\  \label{eq:ampRRQQ} &&{\mathcal
M}_3=2 \pi \alpha_s \frac{q_1^+ q_2^-}{\sqrt{t_1 t_2}}
T^cf^{abc}\frac{\bar U(p_1)\gamma^\mu V(p_2)}{(p_1+p_2)^2}
[(n^-)^\mu(q_2^++\frac{q_2^2}{q_1^-})-(n^+)^\mu(q_1^-+\frac{q_1^2}{q_2^+})+(q_1-q_2)^\mu],
\nonumber
\end{eqnarray}
where $T^a$ are the generators of the fundamental representation of
the color gauge group $SU_C(3)$.

The squared amplitudes can be presented as follows
\begin{eqnarray}
&&\overline{|{\mathcal M}(\mathcal{R}+\mathcal{R} \to c + \bar c)|^2}
= 256 \pi^2 \alpha_s^2 \left( \frac{1}{2 N_c} {\cal A}_{\mathrm{Ab}}
+ \frac{N_c}{2 (N_c^2 - 1)} {\cal A}_{\mathrm{NAb}} \right) \label{sqamp:RRQQ}
\end{eqnarray}

\begin{eqnarray}
&&{\cal A}_{\mathrm{Ab}} = \frac{t_1 t_2}{{\hat t} {\hat u}} -
\left( 1 + \frac{p_2^+}{{\hat
u}}(q_1^--p_2^-)+\frac{p_2^-}{{\hat t}}(q_2^+-p_2^+) \right)^2
\end{eqnarray}

\begin{eqnarray}
{\cal A}_{\mathrm{NAb}} &=& \frac{2}{S^2}\left(\frac{p_2^+
(q_1^--p_2^-)S}{{\hat u}}+\frac{S}{2}+\frac{\Delta}{\hat
s}\right)\left(\frac{p_2^- (q_2^+-p_2^+)S}{{\hat
t}}+\frac{S}{2}-\frac{\Delta}{\hat s}\right)\nonumber\\
&&-\frac{t_1 t_2}{q_1^- q_2^+ {\hat s}}\left(\left(\frac{1}{{\hat
t}}-\frac{1}{{\hat u}}\right)(q_1^- p_2^+ - q_2^+
p_2^-)+\frac{q_1^- q_2^+ {\hat s}}{{\hat t} {\hat u}}-2\right)
\end{eqnarray}

\begin{eqnarray}
&&\Delta = \frac{S}{2}\left({\hat u} - {\hat t}+2 q_1^- p_2^+-2
q_2^+ p_2^- +t_1 \frac{q_2^+-2 p_2^+}{q_2^+}  -t_2 \frac{q_1^--2
p_2^-}{q_1^-} \right)
\end{eqnarray}
Here the bar indicates averaging (summation) over initial-state
(final-state) spins and colors, $t_1 = - q_1^2 = |{\bf q}_{1T}|^2$,
$t_2 = - q_2^2 = |{\bf q}_{2T}|^2$, and
\begin{eqnarray}
&&\hat s = (q_1 + q_2)^2 = (p_1 + p_2)^2\mbox{,}\nonumber\\
&&\hat t = (q_1 - p_1)^2 = (q_2 - p_2)^2\mbox{,}\nonumber\\
&&\hat u = (q_2 - p_1)^2 = (q_1 - p_2)^2\nonumber
\mbox{.}
\end{eqnarray}
The squared amplitude (\ref{sqamp:RRQQ}) analytically coincide with
the previously obtained in Ref.~\cite{KTCollins}. We checked that in
the collinear limit, i.e. $q_{(1,2)T}\to 0$, the squared amplitude
(\ref{sqamp:RRQQ}) after averaging over the azimuthal angles
transforms to the squared amplitude of the corresponding parton
subprocess in collinear model, namely  $g+g\to c+\bar c$. We perform
our analysis in the region of $\sqrt S, p_T\gg m_c$, that allows us
to use zero-mass
 variable-flavor-number-scheme (ZM VFNS), where the masses of the charm  quarks in the hard-scattering amplitude are neglected.

In the $k_T$-factorization, differential cross section for the $2\to
1$ subprocess (\ref{eq:RRg}) has the form:
\begin{eqnarray}
\frac{d \sigma}{dy dp_T}(p + p \to g + X)=  \frac{1}{p_T^3} \int
d\phi_1 \int dt_1 \Phi(x_1,t_1,\mu^2) \Phi(x_2,t_2,\mu^2)
\overline{|{\cal M}(\mathcal{R} + \mathcal{R} \to g)|^2} \mbox{,}
\label{eq:QMRKg}
\end{eqnarray}
where $\phi_1$ is the azimuthal angle between ${\bf p}_T$ and ${\bf
q}_{1T}$.

Analogous formula for the $2\to 2$ subprocess (\ref{eq:RRQQ}) can be
written as
\begin{eqnarray}
&&\frac{d\sigma}{dy_1dy_2dp_{1T}dp_{2T}}(p + p \to c(p_1)+\bar
c(p_2) + X)= \frac{p_{1T}p_{2T}}{16 \pi^3} \int d\phi_1 \int
d\Delta\phi \int dt_1\times \nonumber \\
&& \times \Phi(x_1,t_1,\mu^2) \Phi(x_2,t_2,\mu^2)
\frac{\overline{|{\cal M}(\mathcal{R} + \mathcal{R} \to c + \bar
c)|^2}}{(x_1x_2 S)^2}\mbox{,} \label{eq:QMRKqq}
\end{eqnarray}
where $x_1=q_1^+/P_1^+$, $x_2=q_2^-/P_2^-$, $\Delta\phi$ is the
azimuthal angle between ${\bf p}_{1T}$ and ${\bf p}_{2T}$, the
rapidity of the final-state parton with four-momentum $p$
 is $\displaystyle{y=\frac{1}{2}\ln (\frac{p^+}{p^-})}$.
Again, we have checked a fact that in the limit of $t_{1,2}\to 0$,
we recover the conventional factorization formula of the collinear
parton model from (\ref {eq:QMRKg}) and (\ref{eq:QMRKqq}).

The important ingredient of the our scheme is unintegrated gluon distribution function, which we take as one proposed by Kimber,
  Martin and Ryskin (KMR) \cite{KMR}. These distributions are obtained introducing a single-scale auxiliary function which
  satisfies the unified BFKL/DGLAP evolution equation, where the leading BFKL logarithms $\alpha_S\log (1/x)$ are fully resummed
  and even a major (kinematical) part of the subleading BFKL effects are taken into account.
 This procedure to obtain UPDFs requires less computational efforts than
  the precise solution of two-scale evolution equations such as, for instance, Ciafaloni-Catani-Fiorani-Marchesini
  equation~\cite{CCFM}, but we found it to be suitable and adequate
  to physics of processes under study.

The usage of the $k_T$-factorization formula and UPDFs with one
longitudinal (light-cone) kinematic variable ($x$) requires the
Reggeization of the $t-$channel partons. Accordingly to Refs.~\cite{KTLipatov,LipVyaz},
Reggeized partons carry only one large light-cone component
of the four-momentum and, therefore, it's virtuality is dominated by
the transverse momentum. Such kinematics of the $t-$channel partons
corresponds to the MRK of the initial state radiation and particles,
produced in the hard process. In our previous
analysis~\cite{KniehlSaleevVasin1,KniehlSaleevVasin2,PRD2003,NSS_charm,NSS_bot}
devoted to the similar processes of heavy
  meson production we proved that these UPDFs give the best description of
the heavy quarkonium  $p_T-$spectra measured  at the Tevatron
\cite{CDF} and LHC \cite{LHCdataQuarkonium}.

In the fragmentation model the transition from the produced gluon or
$c$-quark to the $D$-meson is described by fragmentation function
$D_{c,g}(z,\mu^2)$. According to corresponding factorization theorem
of QCD and the fragmentation model, the basic formula for the
$D$-meson production cross section reads \cite{MeleNason}:
\begin{eqnarray}
\frac{d\sigma(p+p\to D+ X)}{dp_{DT} dy}= \sum_i \int_0^1
\frac{dz}{z} D_{i\to D}(z,\mu^2) \frac{d\sigma(p+p\to i(p_i=p_D/z)+
X)}{dp_{iT}dy_i},\label{eq:frag}
\end{eqnarray}
where $D_{i\to D}(z,\mu^2)$ is the fragmentation function for the
parton $i$, produced at the hard scale $\mu$, splitting into
$D$-meson, $z$ is the longitudinal momentum fraction of a
fragmenting particle carried by the $D$-meson. In the zero-mass
approximation the fragmentation parameter $z$ can be defined as
follows $p_D^\mu=zp_i^\mu$, $p_D$ and $p_i$ are the $D-$meson and
$i$-parton four-momenta, and $y_D=y_i$. In our calculations we use
the LO FFs from Ref.~\cite{KniehlKramer}, where the fits
 of non-perturbative $D^0$, $D^+$, $D^{\star +}$, and $D_s^+$ FF's, both at LO and NLO
  in the $\overline{\mbox{MS}}$ factorization scheme, to OPAL data from LEP1~\cite{LEP1} were performed.
 These FFs satisfy two desirable properties: at first, their $\mu-$scaling violation is ruled by DGLAP
  evolution equations; at second, they are universal.

\begin{figure}[ht]
\begin{center}
\includegraphics[width=0.8\textwidth, angle=-90, clip=]{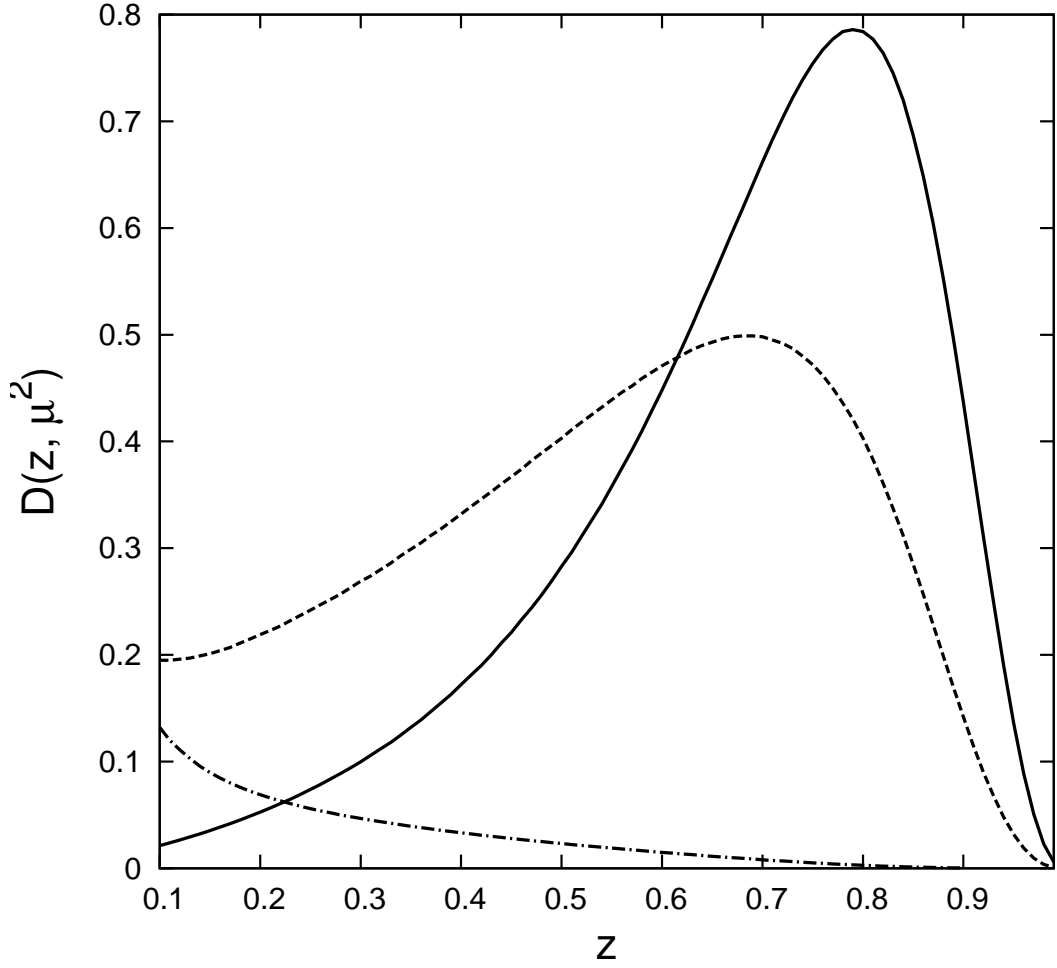}
\end{center}
\caption{The fragmentation function $D(z,\mu^2)$ of $c$-quarks and gluons into $D^{\star}$ mesons from Ref.~\cite{KniehlKramer} at the
 $\mu^2=\mu_0^2=2.25$ GeV$^2$ (solid curve for $c$-quark, the fragmentation function of gluon is negligible) and
 $\mu^2=100$ GeV$^2$ (dashed line for $c$-quark, dash-dotted for gluon).
 \label{fig:dfrag}}
\end{figure}

In the fits of Refs.~\cite{KKSS,KniehlKramer,KKSS2012}, the
parameterizations at the initial scale $\mu_0=m_c$ for the FF's were
taken as follows:
\begin{eqnarray}
&&D_{c}(z,\mu_0^2)=N_{c}\frac{z(1-z)^2}{[(1-z)+\epsilon_{c}]^2}\\
&&D_{g,q}(z,\mu_0^2)=0.\label{eq:start}
\end{eqnarray}
To illustrate a difference of contributions to the $D$-meson production we show in the Fig.~\ref{fig:dfrag}
the $c-$quark and gluon FF's into $D^{\star}-$meson.

As the contribution of gluon fragmentation at $\mu>\mu_0$ is initiated by the perturbative transition of gluons
to $c\bar c$-pairs encountered by DGLAP evolution equations, the part of $c$-quarks produced in the subprocess (\ref{eq:RRQQ})
with their subsequent transition to $D$-mesons are already taken into account considering $D$-meson production via gluon fragmentation.
 Such a way, to avoid double counting, we must subtract this contribution, that can be effectively done
 by the imposing of the lower  cut on $\hat s$ at the threshold of
 the production of the $c\bar c$ pair in (\ref{eq:QMRKqq}), i.e
 $\hat s>4m_c^2$. The precise study of
  double-counting terms and other finite-mass effects needs a separate consideration and can be a subject of our future works.

\section{Results}
\label{sec:three} 
\begin{figure}[ht]
\begin{center}
\includegraphics[width=1.0\textwidth]{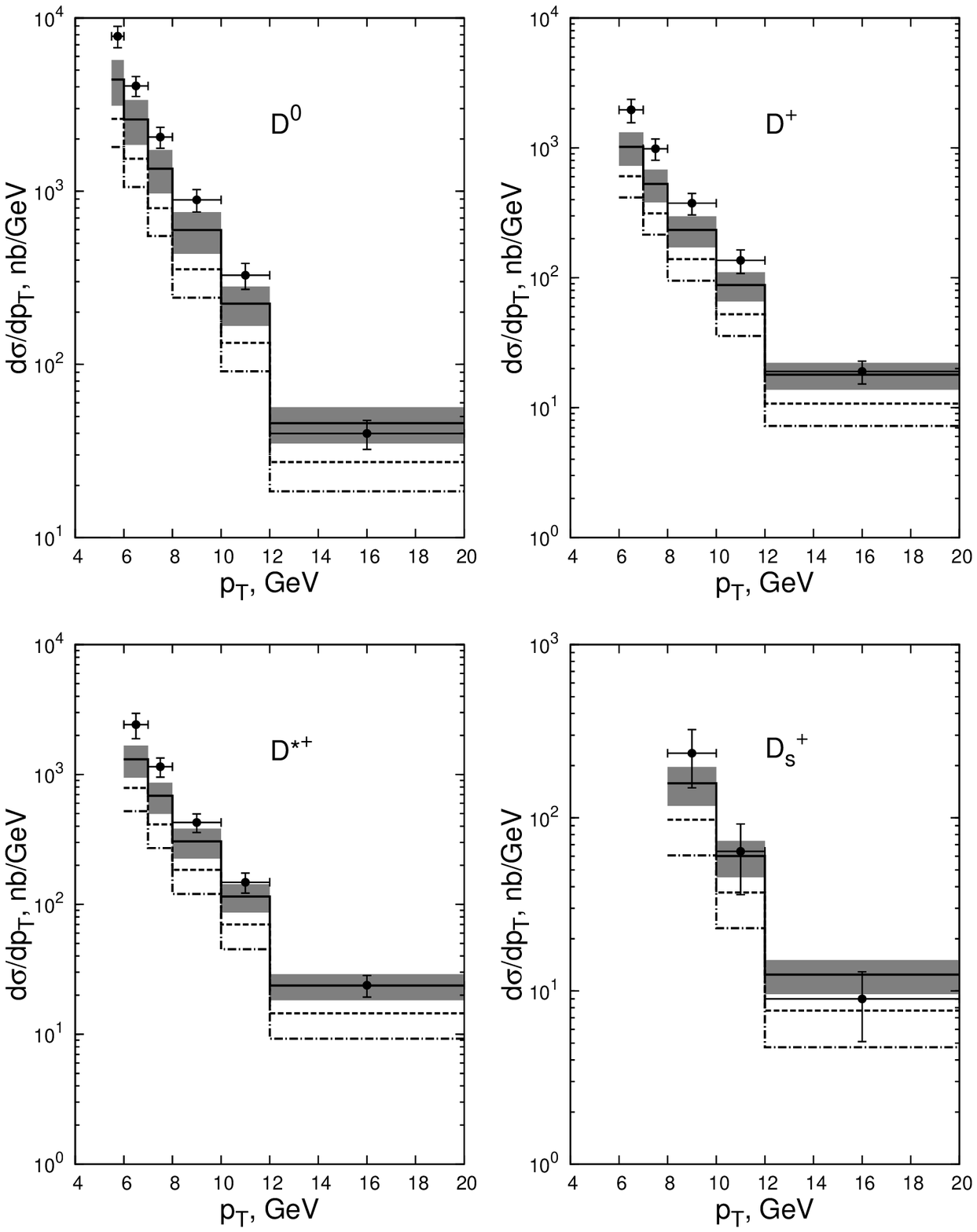}
\end{center}
\caption{Transverse momentum distributions of $D^0$ (left, top), $D^+$ (right, top), $D^{\star +}$ (left, bottom), and $D_s^+$ (right, bottom) mesons
in $p\bar p$ scattering with $\sqrt S=1.96$ TeV and $|y|< 1.0$. Dashed line represents the contribution of gluon fragmentation,
dash-dotted line -- the $c$-quark-fragmentation contribution, solid line is their sum.
The CDF data at Tevatron are from the Ref.~\cite{CDFdmeson}.\label{fig:CDFdmeson}}
\end{figure}

The first measurement of $D$-meson production
transverse-momentum distributions at hadron colliders was
implemented by the CDF Collaboration at Fermilab
Tevatron~\cite{CDFdmeson}, at the collision energy of $\sqrt
S=1.96$~TeV. The production of the $D^0$, $D^+$, $D^{\star+}$, and
$D_s^+$ mesons was studied  in the central region of rapidity
$|y|<1.0$ and  with transverse momenta up to 20~GeV. In the
Fig.~\ref{fig:CDFdmeson} we introduce these data coming as
differential cross sections $d\sigma/dp_T$, where the particle and
antiparticle contributions are averaged, in comparison with our
predictions in the PRA. The dashed lines represent contributions of
the process (\ref{eq:RRg}) while dash-dotted lines correspond to
ones of the process (\ref{eq:RRQQ}). The sum of both contributions
is shown as solid line. We estimated a theoretical uncertainty
arising from uncertainty of definition of factorization and
renormalization scales by varying them between $1/2 m_T$ and $2 m_T$
around their central value of $m_T$, the transverse mass of
fragmenting parton. The resulting uncertainty is depicted in the
figures by shaded bands. We find a good agreement between our
predictions and experimental data in the large-$p_T$ interval of
$D$-meson transverse momenta within experimental
 an theoretical uncertainties.
However, our predictions show a tendency to fall below the data in
the lower $p_T$ range. It can point to the significance of $c$-quark
mass
 effects in the region, where the hard scale of the process is not much larger than the $c$-quark mass.
The increasing of the collision energy with other kinematic conditions preserved is supposed to lead to a better agreement between
theory and experiment in our approach as we expect the rise of logarithmic contributions of type $\log(1/x)$ to be more significant
than finite-quark-mass effects.

\begin{figure}[ph]
\begin{center}
\includegraphics[width=1.0\textwidth]{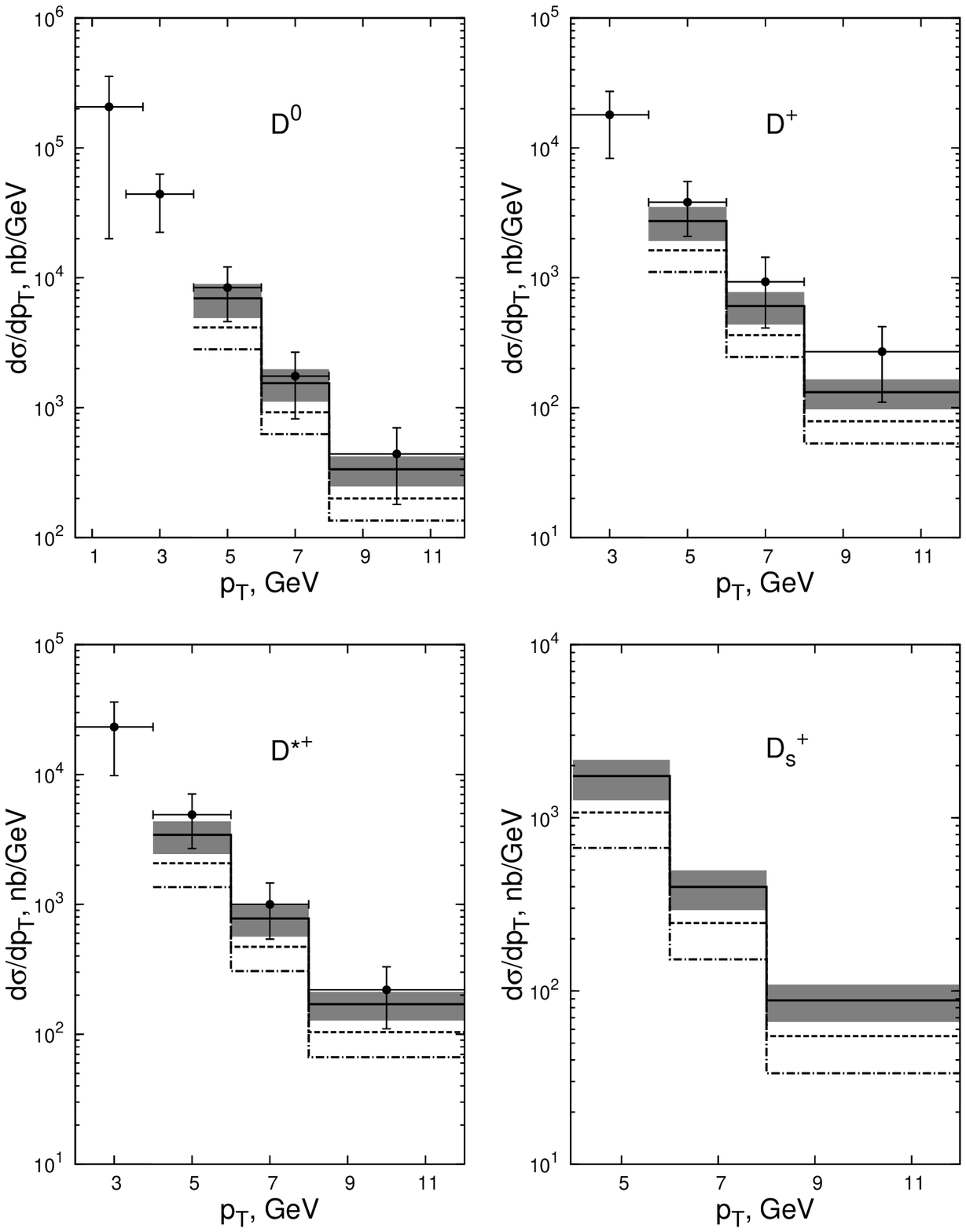}
\caption{Transverse momentum distributions of $D^0$ (left, top), $D^+$ (right, top), $D^{\star +}$ (left, bottom), and $D_s^+$ (right, bottom) mesons
in $pp$ scattering with $\sqrt S=2.76$ TeV and $|y|< 0.5$. The notations as in the Fig.~\ref{fig:CDFdmeson}.
The ALICE data at LHC are from the Ref.~\cite{ALICE276}.~\label{fig:ALICE276}}
\end{center}
\end{figure}

\begin{figure}[ph]
\begin{center}
\includegraphics[width=1.0\textwidth]{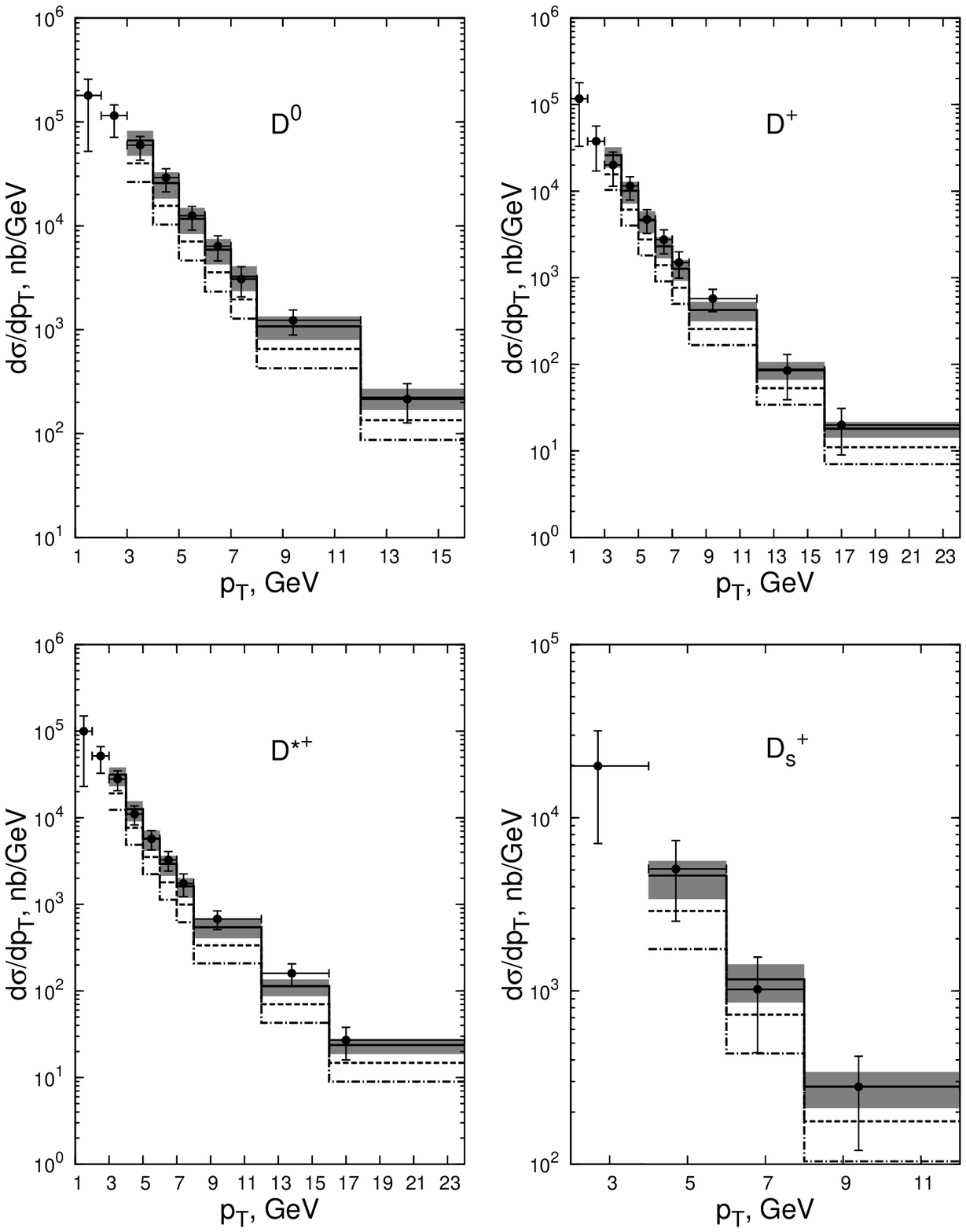}
\caption{Transverse momentum distributions of $D^0$ (left, top), $D^+$ (right, top), $D^{\star +}$ (left, bottom), and $D_s^+$ (right, bottom) mesons
in $pp$ scattering with $\sqrt S=7$ TeV and $|y|< 0.5$. The notations as in the Fig.~\ref{fig:CDFdmeson}.
The ALICE data at LHC are from the Ref.~\cite{ALICE7}. \label{fig:ALICE7}.}
\end{center}
\end{figure}

\begin{figure}[ht]
\begin{center}
\includegraphics[width=1.0\textwidth]{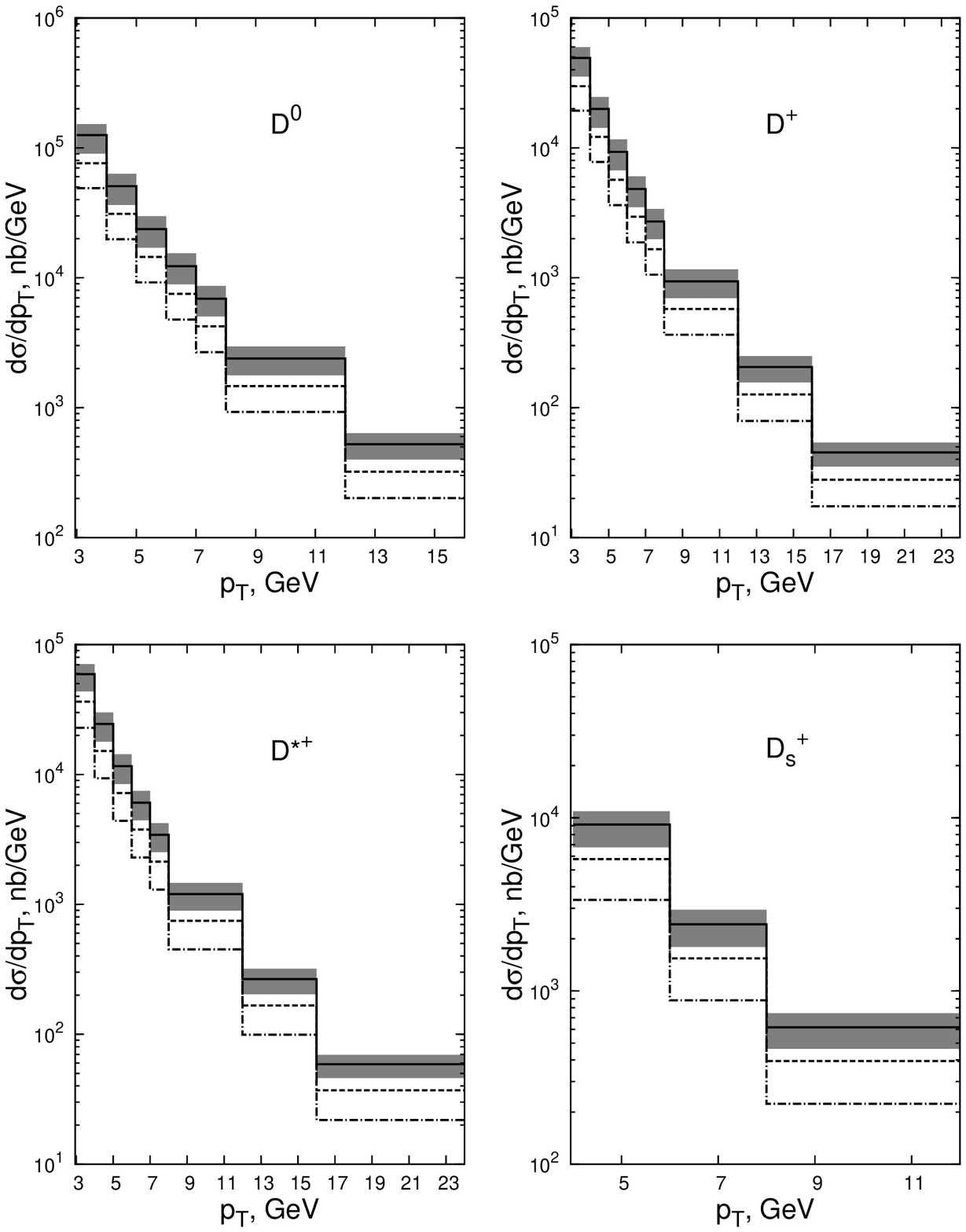}
\end{center}
\caption{Theoretical predictions for the transverse momentum distributions of $D^0$ (left, top), $D^+$ (right, top), $D^{\star +}$ (left, bottom), and $D_s^+$ (right, bottom) mesons
in $pp$ scattering at $\sqrt S=14$~TeV and $|y|< 0.5$ obtained in the LO PRA. The notations as in the Fig.~\ref{fig:CDFdmeson}.
 \label{fig:14}}
\end{figure}

Our expectations are confirmed when we turn to the description of
the recent data from the LHC at  its intermediate energy of $\sqrt
S=2.76$~TeV and $\sqrt S=7$~TeV collected by the ALICE
Collaboration~\cite{ALICE276,ALICE7}. The previous NLO predictions
made in collinear parton model in general are in agreement with
ALICE data, however, one can find that the FONLL
scheme~\cite{newNLO} tend to overestimate data and the
GM-VFN~\cite{KKSS2012} is to underestimate. In the
Figs.~\ref{fig:ALICE276} and~\ref{fig:ALICE7}, we compare our
predictions with ALICE data~\cite{ALICE276,ALICE7} keeping the
notations of curves the same as in the Fig.~\ref{fig:CDFdmeson}. The
current collision energies of LHC is 2-3.5 times larger compared to
Tevatron and the interval of $D$-meson rapidity is more narrow,
$|y|<0.5$. We obtain a good agreement  of our predictions with the
experiment for the all types of $D$-mesons at the whole range of
their transverse momenta. As there is no experimental data for
$D_s^+$-production at the energy of $\sqrt S=2.76$~TeV, we introduce
the theoretical prediction only. Finally, in the Fig.~\ref{fig:14}
we present our predictions for the planned LHC energy of $\sqrt
S=14$~TeV and the other kinematic conditions as in the
Ref.~\cite{ALICE7}.

Considering the $D$-meson central rapidity production, we find the
MRK subprocess (\ref{eq:RRg}) to remain indeed the dominant one for
all collision energies. Such a way, we confirm the theoretical
suggestion mentioned in Sec.~\ref{sec:two} that MRK gives the
leading logarithmic approximation for the high-energy production
processes in BFKL approach while the QMRK turns out to be
subleading. However, in the framework of Ref.~\cite{newKT}, which
seems to be theoretically close to PRA, this MRK subprocess is
absent while the main contribution is coming from QMRK subprocess.

\begin{figure}[ph]
\begin{center}
\includegraphics[width=1.0\textwidth]{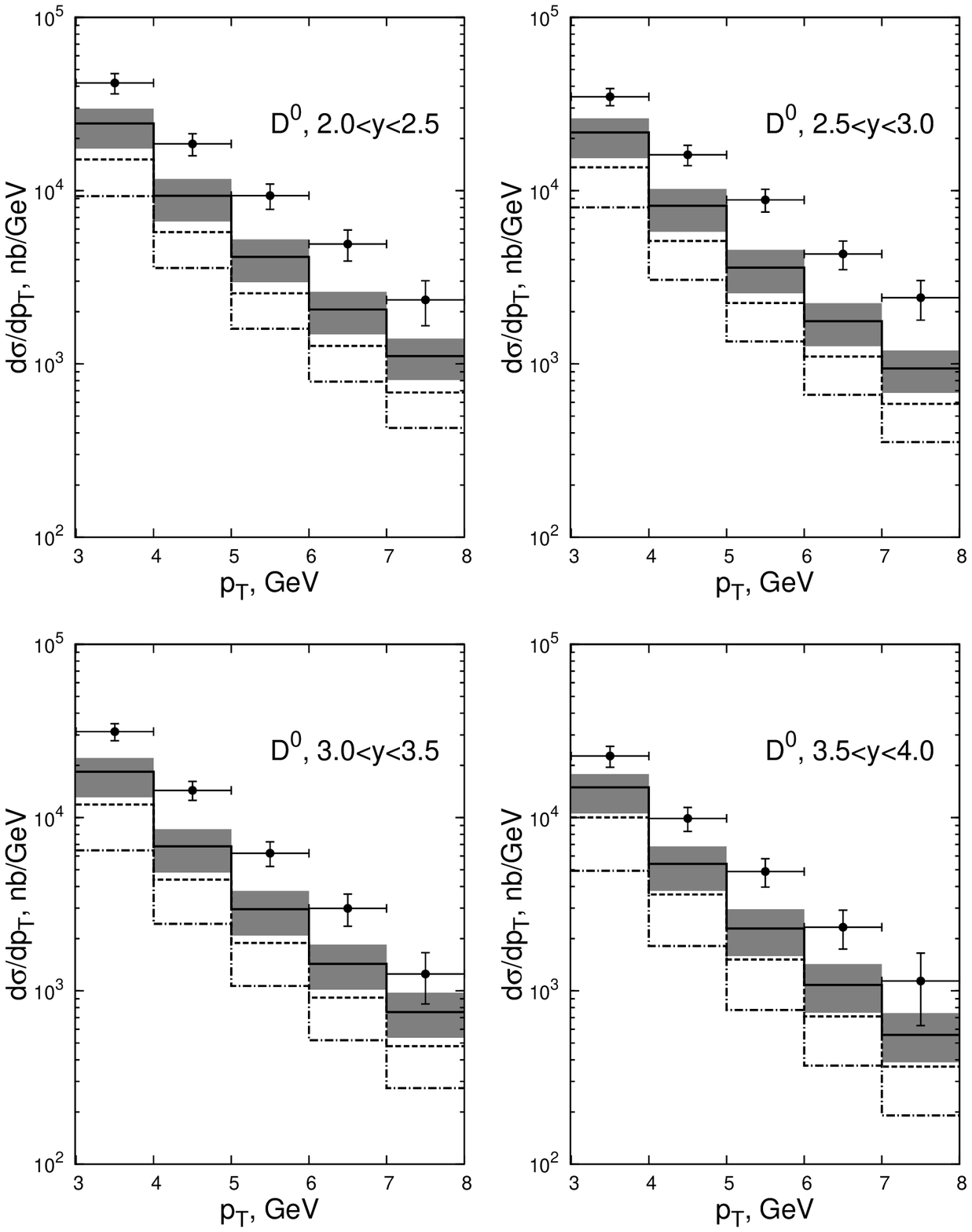}
\caption{ Transverse momentum distributions of $D^0$ mesons in forward rapidity region
in $pp$ scattering with $\sqrt S=7$ TeV. The LHCb data at LHC are from the Ref.~\cite{forward}. The notations as in the Fig.~\ref{fig:CDFdmeson}.\label{fig:BRD0}}
\end{center}
\end{figure}

\begin{figure}[ph]
\begin{center}
\includegraphics[width=1.0\textwidth]{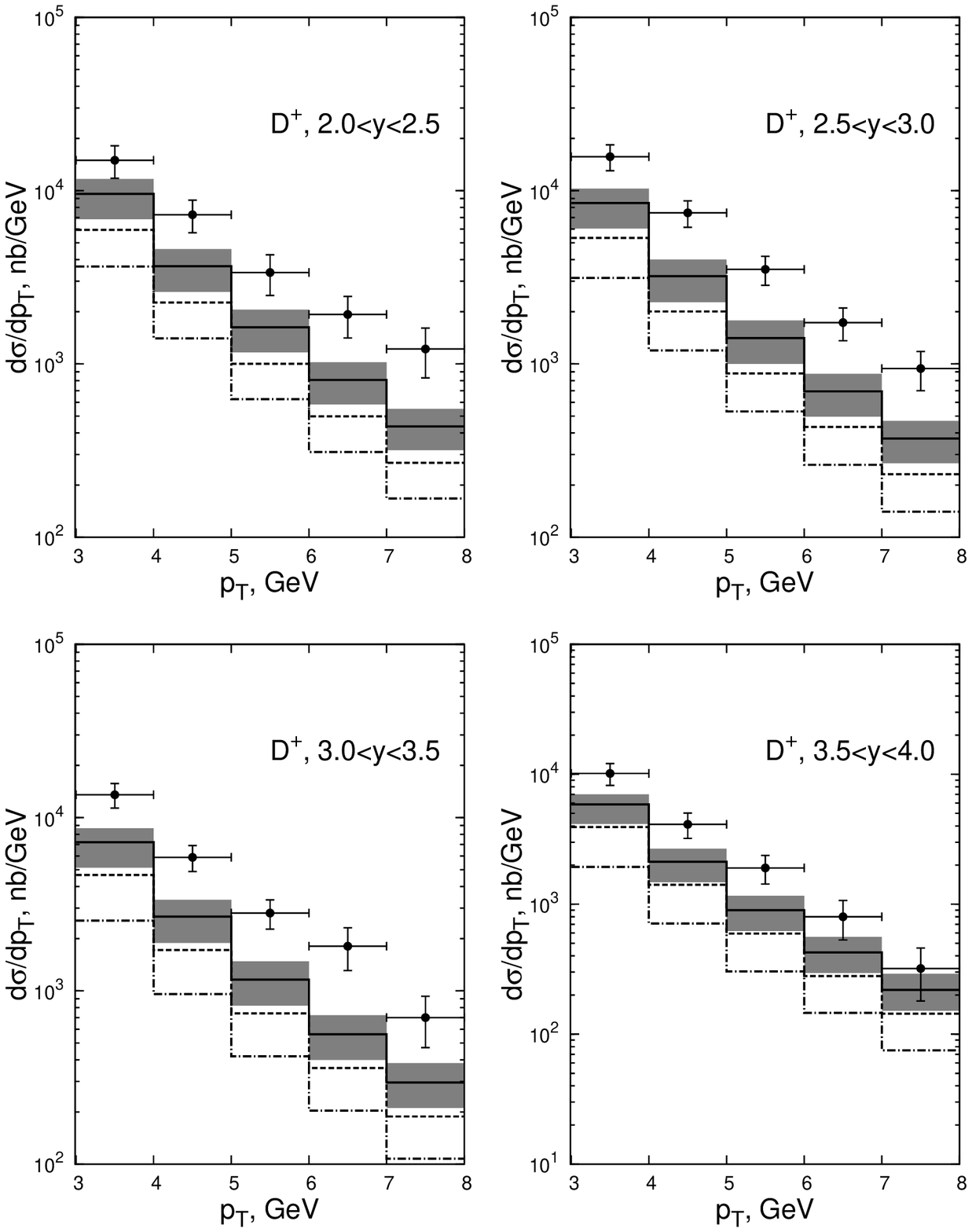}
\caption{ The same as in the Fig.~\ref{fig:BRD0} for $D^+$ mesons}.\label{fig:BRDplus}
\end{center}
\end{figure}

\begin{figure}[ph]
\begin{center}
\includegraphics[width=1.0\textwidth]{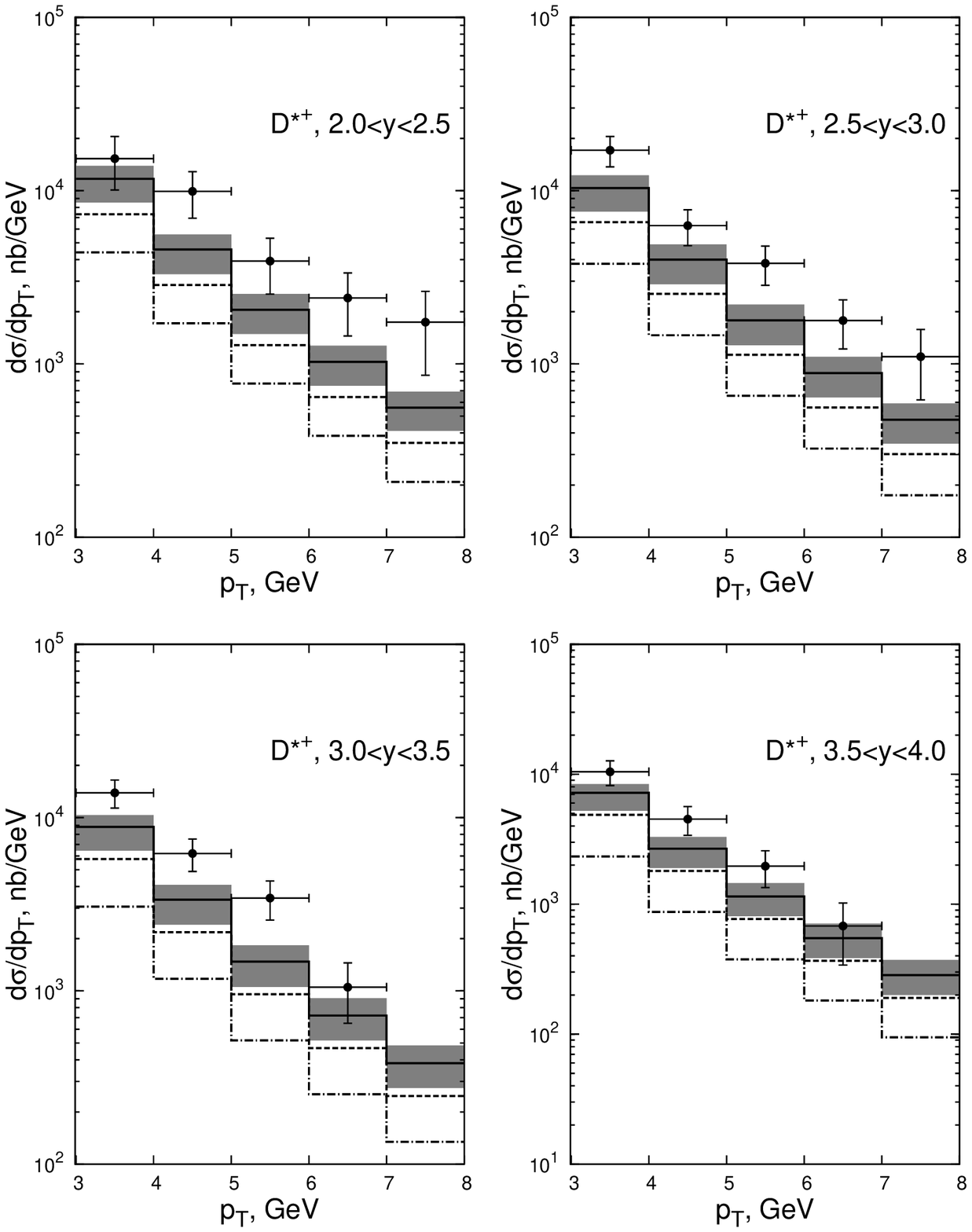}
\caption{ The same as in the Fig.~\ref{fig:BRD0} for $D^{\star +}$ mesons. \label{fig:BRDstar}}
\end{center}
\end{figure}

\begin{figure}[ph]
\begin{center}
\includegraphics[width=1.0\textwidth]{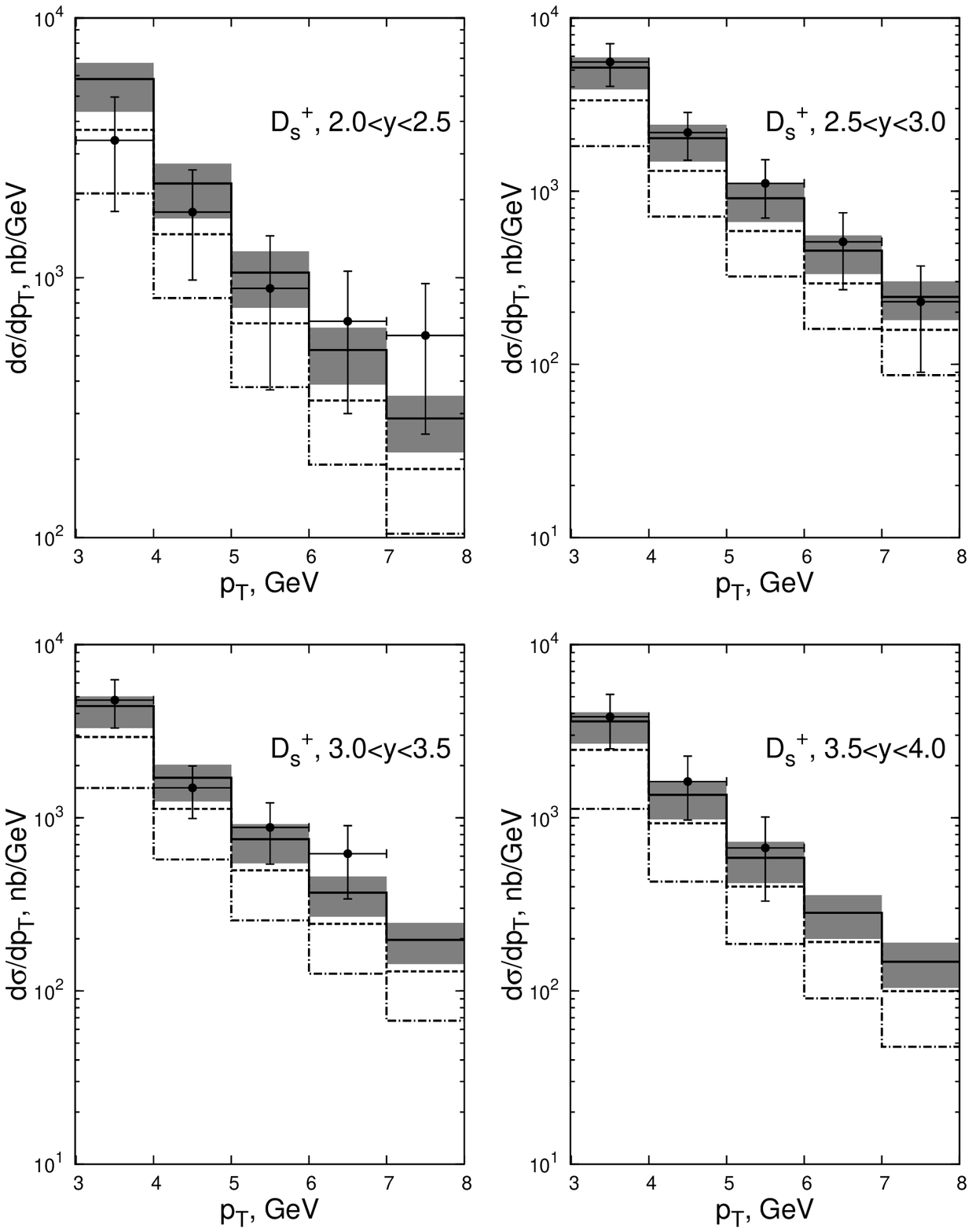}
\caption{ The same as in the Fig.~\ref{fig:BRD0} for $D_s^+$ mesons.\label{fig:BRDs}}
\end{center}
\end{figure}

\begin{figure}[ph]
\begin{center}
\includegraphics[width=1.0\textwidth]{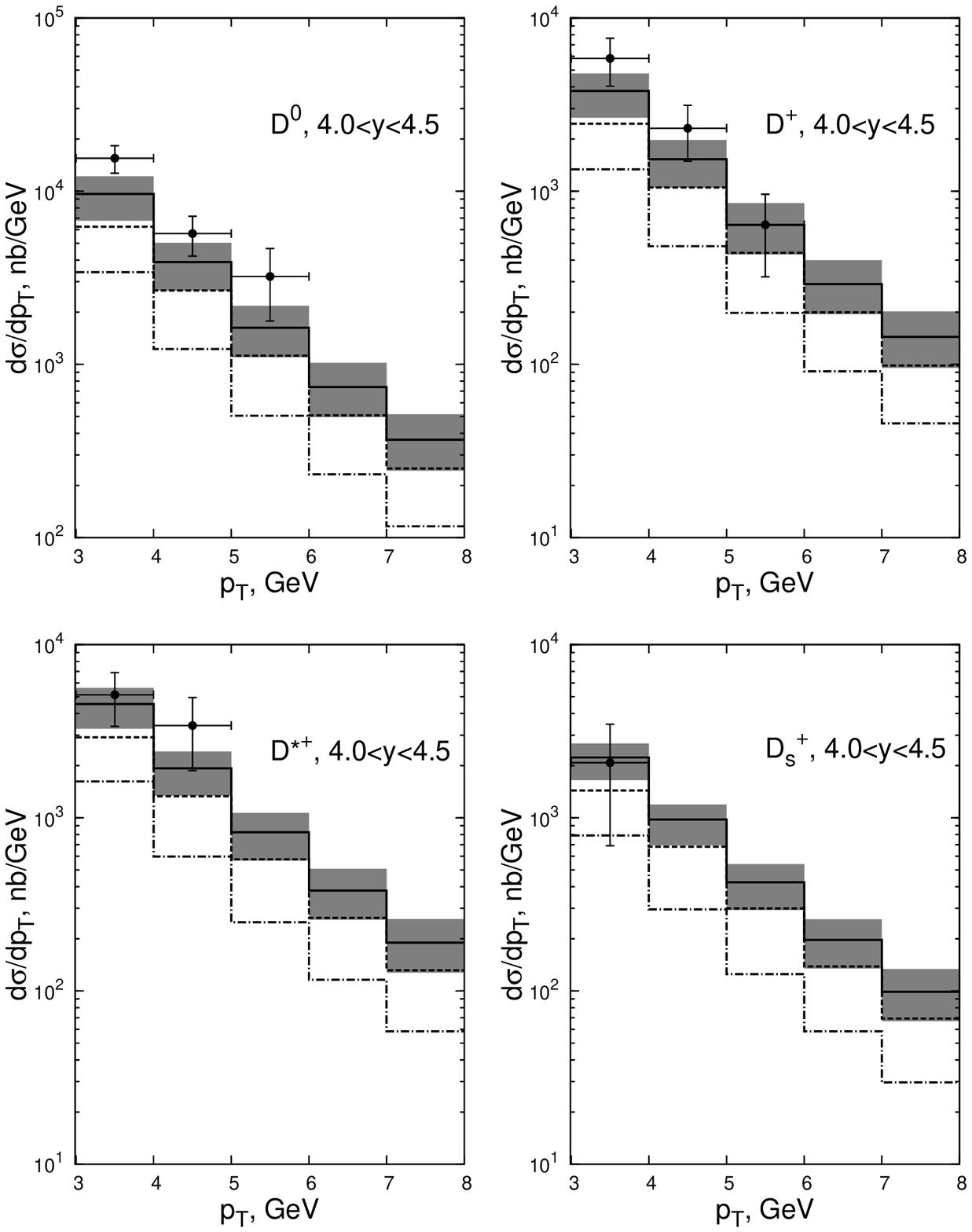}
\caption{ Transverse momentum distributions of $D^0$ (left, top), $D^+$ (right, top), $D^{\star +}$ (left, bottom), and $D_s^+$ (right, bottom) mesons in forward rapidity region
in $pp$ scattering with $\sqrt S=7$ TeV and $4.0<y<4.5$. The LHCb data at LHC are from the Ref.~\cite{forward}. The notations as in the Fig.~\ref{fig:CDFdmeson}.\label{fig:BR5}}
\end{center}
\end{figure}

Not only the central but also the forward rapidity region in $pp$
collisions at the LHC become available due to the specially designed
LHCb detector where the measurements of differential cross sections
of $D^0$, $D^+$, $D^{\star+}$, and $D_s^+$ mesons with $2.0<y<4.5$
at $\sqrt S=7$~TeV were performed~\cite{forward}. The observed data
divided into 5 rapidity regions have been under study in both, FONLL
and GM-VFN, schemes and were founded to be generally enclosed
between their predictions~\cite{KKSS2012, newNLO}. We present these
data together with our results obtained in the LO of PRA in the
Figs.~\ref{fig:BRD0}-\ref{fig:BR5}. One can find the summary
contribution mainly to underestimate the data from 1.5 to 2 times
with a slight exception in the case of $D_s^+$ production. This
result is expected and becomes clear if we recall that with grow of
rapidity of the particle produced in the hard scattering process the
fraction of longitudinal momenta of initial proton transferred to
this process increases simultaneously. That means, on a one hand,
that we enter the region of large $x>0.1$ where the conditions of
Reggeization are not satisfied, and another effects, such as signals
of intrinsic charm, can appear. On the other hand, to balance the
large positive rapidity of a produced particle one needs a very
small negative longitudinal momenta to income the hard subprocess
from the side of another proton in the collision. Such a way, we
have a very asymmetric case where the one $t$-channel exchange is
perfectly under the multi-Regge kinematics conditions being strong
opposite the second one. That leads to the situation in which we
finely take into account small-$x$ effects although loosing in
large-$x$. It is illustrated by the Fig.~\ref{fig:BR5} dedicated to
the largest rapidity region $4.0<y<4.5$ where we obtain a better
agreement with experimental data in comparison with other ones of
forward production. It proves our assumption that BFKL-type
logarithms exhibit themselves at the already achieved collision
energies giving a significant contribution to the production rates.

Considering the relative contributions of the subprocesses in the forward rapidity region, we find the QMRK subprocess to decrease with grow
 of rapidity. That illustrates the fact that the probability of a production of quark and antiquark both with large close rapidities in a
 symmetric collision diminishes. The case when one of them has a significant positive rapidity, and another one -- the same negative,
 contradicts the definition of QMRK process.

\section{Conclusions}
\label{sec:four}

We introduce a comprehensive study of  $D^0$,
$D^+$, $D^{\star +}$, and $D_s^+$-meson fragmentation production in proton-(anti)proton collisions with central rapidities
at Tevatron Collider and LHC and in the forward rapidity region for the LHC, in the framework of Parton Reggeization Approach.
We use the gauge invariant amplitudes of hard parton subprocesses in the LO level of theory with Reggeized gluons in the initial state
 in a self-consistent way together with unintegrated parton distribution functions proposed by Kimber, Martin and Ryskin.
 The $2 \to 1$ hard subprocess of gluon production via a fusion of two Reggeized gluons in the PRA framework is proposed for
 the first time in the case of $D$-meson fragmentation production and proved to be a dominant one. To describe the non-perturbative
 transition of produced gluons and $c$-quarks into the $D$-mesons we use the universal fragmentation functions obtained from the fit
  of $e^+ e^-$ annihilation data from CERN LEP1.
 We found our results for $D$-meson central-rapidity production to be in the excellent coincidence with experimental data from the LHC
  and good agreement with large-transverse-momenta Tevatron data. The achieved degree of agreement for the LHC exceeds the one obtained
   by NLO calculations in
  the conventional collinear parton model and LO calculations in $k_T$-factorization with Reggeized gluons. The predictions for the
  $D$-meson production in the central rapidity region for the expected LHC energy of $\sqrt S=14$ TeV are also presented.
  For the forward rapidity region we compare our results with transverse-momentum $D$-meson distributions measured by LHCb Collaboration at LHC,
  and the expected discrepancies are obtained. We describe
  $D$-meson production without any free parameters or auxiliary approximations.

\section{Acknowledgements}
\label{sec:five} The work of A.~V.~Shipilova and A.~V.~Karpishkov
was partly supported by the Grant of President of Russian Federation
No. MK-4150.2014.2. The work of M.A.~Nefedov and V.A.~Saleev  was
supported in part by the Russian Foundation for Basic Research
through the Grant No. 14-02-00021. A.~V.~Shipilova is grateful
to Prof. G.~Kramer for the useful discussions, to Prof. B.~A.~Kniehl for the kind hospitality,
and to the German Academic Exchange Service (DAAD) together with the Russian Federal Ministry of Science and Education
for the financial support by Grant~No.~A/13/75500.

\newpage

\newpage

\newpage

\newpage

\newpage

\newpage

\newpage

\newpage

\newpage

\newpage

\newpage

\end{document}